\documentstyle[11pt,amsfonts]{article}

\newcommand{\be}{\begin{equation}}
\newcommand{\ee}{\end{equation}}
\newcommand{\bea}{\begin{array}}
\newcommand{\ea}{\end{array}}
\newcommand{\beqa}{\begin{eqnarray}}
\newcommand{\eeqa}{\end{eqnarray}}
\newcommand{\bean}{\begin{eqnarray*}}
\newcommand{\eean}{\end{eqnarray*}}

\def\up#1{\leavevmode \raise.16ex\hbox{#1}}

\newcommand{\gapproxeq}{\lower
 .7ex\hbox{$\;\stackrel{\textstyle >}{\sim}\;$}}
\newcommand{\lapproxeq}{\lower .7ex\hbox{$\;\stackrel
{\textstyle <}{\sim}\;$}}

\newcounter{appendice}

\def\thebibliography#1{{\bf REFERENCES\markboth
 {REFERENCES}{REFERENCES}}\list
 {[\arabic{enumi}]}{\settowidth\labelwidth{[#1]}\leftmargin\labelwidth
 \advance\leftmargin\labelsep
 \usecounter{enumi}}
 \def\newblock{\hskip .11em plus .33em minus -.07em}
 \sloppy
 \sfcode`\.=1000\relax}

\def\BI{{\rm 1\!l}}

\begin{document}
\vskip 1cm
\centerline{ \LARGE Relativistic Particles on Quantum Space-time }

\vskip 2cm

\centerline{ {\sc  A. Stern }\footnote{astern@bama.ua.edu}   }

\vskip 5mm

\centerline{  Dept. of Physics and Astronomy, Univ. of Alabama,
Tuscaloosa, Al 35487, U.S.A.}
\vskip 2cm

\vspace*{5mm}

\normalsize
\centerline{\bf ABSTRACT}

We discuss alternatives to the usual quantization of 
a  relativistic particle which  result in   discrete spectra for  position and time operators.

\vskip 4 cm

\section{Introduction}
\setcounter{equation}{0}

Space-time noncommutativity can arise from the quantization of a point particle.\cite{Pinzul:2004qu},\cite{Banerjee:2004ms}
   For this one only needs to  start with a reparametrization invariant action and then employ alternatives to the standard gauge fixing condition that identifies the particle's  world line parameter  with the time-component of the position four-vector.  Alternative gauge fixing  conditions can lead to nontrivial Dirac brackets between different components of  the position four-vector, and  noncommutative space-time appears upon replacing the Dirac brackets with quantum commutators. 
   The procedure has been employed in order to recover various interesting deformations of the Heisenberg algebra.\cite{Romero:2004er},\cite{Ghosh:2005rz},\cite{Girelli:2005dc},\cite{Malik:2005de},
\cite{Banerjee:2006wf},\cite{Gangopadhyay:2007rh} In particular, for a suitable gauge choice one obtains  the Snyder algebra,\cite{Snyder:1946qz} which is a Lorentz covariant deformation and   is characterized by a discrete spectrum for 
 the position operators.  With another gauge choice, a 
particle action written on a continuous space-time can lead to a  discrete spectrum for the time operator in the quantum theory.\cite{Romero:2007hi}
The quantum description of  the particle in these gauges is thus distinct from what one obtains in the standard gauge.  This gauge dependence for the quantum description of the particle is analogous to the presence of anomalies in quantum field theory.

 The previous derivations of Snyder's algebra and its resulting discretized postion spectra have involved either deforming the standard action for a relativistic free particle   or introducing extra degrees of freedom to the system, which are later eliminated  using  gauge fixing conditions.\cite{Romero:2004er},\cite{Banerjee:2006wf},\cite{Romero:2006pe},\cite{Jaroszkiewicz}  By the standard action we are referring to 
 \be
S =-m\int d\lambda \;\sqrt{-\dot x_\mu \dot x^\mu}\;, \label{fpa}
\ee 
or equivalent reparametrization invariant expressions describing  a massive particle.  Here
 $x^\mu(\lambda)$, $\mu=0,1,...3$, defines the trajectory of  the particle, $m$ is the mass,
$\lambda$ is an arbitrary evolution parameter and 
 the dot indicates differentiation
with respect to $\lambda$.   We choose  $c=1$ and metric tensor $[g_{\mu\nu}]=$diag$(-1,1,1,1)$.
In this article we show that a  discrete spectra for the position operators can be obtained  directly from (\ref{fpa}), {\it without deforming  the action
or introducing additional degrees of freedom to the system}.   This only requires finding a suitable gauge condition to fix $\lambda$.   We do not recover the full Lorentz covariant algebra of Snyder in this case because, like with the standard gauge,  the gauge condition breaks Lorentz covariance. Nevertheless, the  algebra we obtain is  sufficient for getting a discretized space in the quantum theory, and the full Poincar\'e algebra is    realized by the Dirac brackets.  The  quantum theory carries the spin zero irreducible representation of the Poincar\'e group, and, like with the standard gauge, the zero-component of the four-momentum serves as a Hamiltonian for the theory,  generating evolution along the trajectory.

Starting with yet another gauge fixing of the parameter $\lambda $ in  (\ref{fpa}), one can get a discrete spectrum for the time operator.  Unlike in \cite{Romero:2007hi}, we shall not  introduce  additional particle degrees of freedom for this purpose.  Instead, we shall only require that one of the spatial coordinates $x^i$ in  (\ref{fpa}) 
 be an angular variable, thereby implying the existence of  a coordinate singularity.  Alternatively, the latter can be promoted to a real singularity by replacing $g_{\mu\nu}$ by a black hole metric. The resulting quantum algebra agrees  with what was found previously for the BTZ black hole from symmetry considerations,\cite{Dolan:2006hv}  and aspects of the quantum theory were  studied previously by  several authors.\cite{Chaichian:2001nw},\cite{Balachandran:2004yh}  

After first reviewing the standard gauge in section two, we give the gauge condition which leads to discretized space in section three, and discretized time in section four.  
 \bigskip

\section{Standard  gauge }
\setcounter{equation}{0}
Using the Dirac Hamiltonian formalism,\cite{Rht} we now review the standard gauge constraint.   
  The
 four-momenta, 
\be p_\mu = \frac{m\dot
  x_\mu}{\sqrt{- \dot x^\nu \dot x_\nu}}\;, \label{smlpmu}\ee 
 obtained from (\ref{fpa})  are canonically
conjugate to the space-time coordinates, 
\be 
\{x^\mu,p_\nu\} =\delta^\mu_\nu \qquad \{x^\mu,x^\nu\} =
\{p_\mu,p_\nu\} =0\label{cpb}\;, \ee
 and are subject to  the mass shell condition
\be
\Psi_1 =p^\mu p_\mu + m^2 \approx 0\;,\label{mschl}
\ee  where $\approx$   indicates equality  in weak sense. $\Psi_1$ generates gauge motion on the phase space associated with
reparametrizations of $\lambda$.  
The gauge symmetry is  fixed after imposing an additional constraint $\Psi_2\approx 0$.  The standard choice for $\Psi_2$ identifies $\lambda$ with  the time coordinate $x^0$,
\be \Psi_2 = x^0  - \lambda\;\approx 0\label{stndrdgfxng} \;,\ee
and, as a result, (\ref{mschl}) and (\ref{stndrdgfxng}) form a  second class
set.  Dirac brackets\cite{Rht} 
\be \{A,B\}_{\mbox{\tiny DB}} = \{A,B\} + \frac1{ \{\Psi_1,\Psi_2\}} \biggl(
\{A,\Psi_1\}\{\Psi_2,B\} -\{A,\Psi_2\}\{\Psi_1,B\}\biggr)\label{DB}\;, \ee
are then employed to write down a consistent algebra on  phase space.
The result is
\be\{x^\mu,p_\nu\}_{\mbox{\tiny DB}} = \delta^\mu_\nu - \delta^0_\nu \;\frac {p^\mu}{p^0} \qquad\quad \{x^\mu,x^\nu\}_{\mbox{\tiny DB}} = \{p_\mu,p_\nu\}_{\mbox{\tiny DB}} =0\;\label{stndrdgagdbs}\ee
It is not Lorentz covariant, since neither is (\ref{stndrdgfxng}).  The Poincar\'e algebra easily follows from the Dirac brackets (\ref{stndrdgagdbs}), 
\beqa \{j^{\mu\nu},p^\rho\}_{\mbox{\tiny DB}}&=&\eta^{\mu\rho} p^\nu-\eta^{\nu\rho} p^\mu \label{poincarealgpj}\\& &\cr 
 \{j^{\mu\nu},j^{\rho\sigma}\}_{\mbox{\tiny DB}}&=&\eta^{\mu\rho}j^{\nu\sigma} -\eta^{\nu\rho}j^{\mu\sigma}-\eta^{\mu\sigma}j^{\nu\rho}+\eta^{\nu\sigma}j^{\mu\rho}\label{poincarealgjj}\;,\eeqa
 where \be j^{\mu\nu}= x^\mu p^\nu -  x^\nu p^\mu  \;,\label{jpdbs}\ee
and so the quantum theory carries the spinless irreducible representation of the Poincar\'e group.  The quantum  operators $ \hat x^i$, associated with the position coordinates $ x^i$, have  continuous spectra, while the time coordinate  $x^0$ remains a commuting parameter in the quantum theory. 
 $p^0$ serves as the Hamiltonian for the system, generating evolution in $x^0$, i.e.,  for any function ${\cal F}(x,p,\lambda) $ on the phase space, 
\be \frac{d}{dx^0}{\cal F}(x,p)  = \{{\cal F}(x,p) ,{\cal H}\}_{\mbox{\tiny DB}}+   \frac{\partial }{\partial x^0}{\cal F}(x,p)\;,\qquad {\cal H}=p^0\; \label{2.10}\ee
The Hamilton equation  (\ref{2.10}) gets replaced by the corresponding Heisenberg equation in the quantum theory.

\section{Discrete space gauge}
\setcounter{equation}{0}

We now introduce an alternative to the  gauge fixing condition (\ref{stndrdgfxng}), which  deforms the Dirac brackets (\ref{stndrdgagdbs}), while preserving the Poincar\'e algebra (\ref{poincarealgpj}),(\ref{poincarealgjj}).  It  leads to a different quantum description of the relativistic free particle.  The alternative gauge condition is\footnote{ It is similar to a gauge condition used in \cite{Romero:2006pe}.} 
\be \Psi_2 = x^0 +\frac{p_0\;\vec x\cdot \vec p}{\Lambda^2 +\vec p^2} - \lambda\;\approx 0\label{gfxng} \;,\ee
where $\vec x$ and $\vec p$ denote three-vectors and $\Lambda$ is some energy scale ($\Lambda\ne 0$).  The new term in $\Psi_2$ vanishes in the limit $\Lambda\rightarrow \infty$ and so we recover the standard  gauge fixing condition in the limit. The gauge condition (\ref{gfxng}) says that the evolution parameter $\lambda$ is a momentum dependent rescaling  of the time coordinate  $x^0$ along any free particle world line, \be\frac{d\lambda}{dx^0}\Big|_{p_\mu={\rm const}}=\frac{\Lambda^2}{\Lambda^2+ \vec p^2}\label{dlmbddt}\ee [To obtain this result set (\ref{gfxng}) strongly equal to zero and use ${dx^i}/{dx^0} = {\dot x^i}/{\dot x^0}={p^i}/{p^0}$, which is valid along the particle world line.]   Then $\lambda$  increases monotonically as the particle evolves in the time-like direction, and  (\ref{gfxng}) a valid gauge condition.

 From (\ref{DB}), one now gets the following Dirac brackets 
\beqa \{x_i,x_j\}_{\mbox{\tiny DB}} &=&\frac 1{\Lambda^2}\epsilon_{ijk}L_k\label{xixjdb}\\& & \cr \{x_i,p_j\}_{\mbox{\tiny DB}} &=&\delta_{ij} +\frac{p_ip_j}{\Lambda^2}\\& &\cr \{p_i,p_j\}_{\mbox{\tiny DB}} &=&0\label{snydrsubalg}\eeqa
where $i,j,k=1,2,3$ and $L_i=\epsilon_{ijk}x_j p_k$ is the angular momentum.  This is the classical analogue of Snyder's algebra
restricted to the reduced phase space spanned by $x^i$ and $p_i$.  We do not recover the full Lorentz covariant algebra of Snyder because, as was true in section two, the gauge constraint spoils Lorentz covariance.  
   For  the time-like components $x^0$ and $p^0$, one instead gets the Dirac brackets
\beqa \{x^0,x_i\}_{\mbox{\tiny DB}}&=&\frac 1{\Lambda^2}\biggl(p_0 x_i +\frac{\vec x\cdot \vec p (\Lambda^2 - p_0^2-m^2)}{p_0(\Lambda^2 +\vec p^2)}\;p_i\biggr)\\& &\cr
\{x_i,p_0\}_{\mbox{\tiny DB}} &=&\Bigl(1+\frac{\vec p^2}{\Lambda^2}\Bigr) \frac {p_i}{p_0}\label{xip0}\\& &\cr
 \{x_0,p_0\}_{\mbox{\tiny DB}}&=&\frac{\vec p^2}{\Lambda^2}\label{x0p0}\\& &\cr
  \{x_0,p_i\}_{\mbox{\tiny DB}}&=&\frac{p_0p_i}{\Lambda^2}\;,\label{DBof xipj}\eeqa along with
 $\{p_0,p_j\}_{\mbox{\tiny DB}} =0$. The Dirac brackets  (\ref{xixjdb}-\ref{DBof xipj}) reduce to those of the standard gauge  (\ref{stndrdgagdbs}) in the limit $\Lambda\rightarrow\infty$.
 
 Using  the definition   
 (\ref{jpdbs}) of the Lorentz generators and the Dirac brackets  (\ref{xixjdb}-\ref{DBof xipj}), it can be checked that (\ref{poincarealgpj}) and (\ref{poincarealgjj}) are satisfied for all $\Lambda\ne 0$.  The Poincar\'e algebra is therefore recovered and, like in the standard gauge, the quantum theory carries the spinless irreducible representation  of the Poincar\'e group.   
Also, as  in the standard gauge, $p^0$ serves as the Hamiltonian for the system.  Here, though, it generates evolution in $\lambda$, and not $x^0$.
 From (\ref{dlmbddt}), (\ref{xip0}) and (\ref{x0p0}), one gets
\be  \{x^\mu,p^0\}_{\mbox{\tiny DB}}+   \frac{\partial x^\mu}{\partial\lambda}=\frac{\frac{dx^\mu}{dx^0}} {\frac{d\lambda}{dx^0}} =\frac{dx^\mu}{d\lambda}\;,\ee
where $x_i$, $p_i$ and $\lambda$ are regarded as independent variables in the partial derivative, while $x^0$ is  defined using the  constraint  (\ref{gfxng}), and so $ \frac{\partial x^\mu}{\partial\lambda}=\delta^\mu_0$. 
 Then for any function ${\cal F}(x,p,\lambda) $ on the phase space, one has the  Hamilton equation
\be \frac{d}{d\lambda}{\cal F}(x,p,\lambda)  = \{{\cal F}(x,p,\lambda) ,{\cal H}\}_{\mbox{\tiny DB}}+   \frac{\partial }{\partial\lambda}{\cal F}(x,p,\lambda)\;,\qquad {\cal H}=p^0\;, \label{Hameq}\ee
To obtain the classical evolution in $x^0$, one can first solve (\ref{Hameq}) and then apply (\ref{dlmbddt}).\footnote{Alternatively, we can search for the generator ${\cal P}_0$ of translations in the time coordinate $x^0$. On the reduced phase space spanned by $x^i$ and $p_i$, it should satisfy the Dirac brackets
 $$\{x^i,{\cal P}_{0}\}_{\mbox{\tiny DB}}=\frac{dx^i}{dx^0}\quad\qquad\{p_i,{\cal P}_{0}\}_{\mbox{\tiny DB}}=0$$   
A solution  is $${\cal P}_{0}= \frac{\Lambda^2}{\sqrt{\Lambda^2-m^2}}\;\tan^{-1}\frac{p^0}{\sqrt{\Lambda^2-m^2}}\;,$$   which is valid  for $\Lambda > m$.
 ${\cal P}_{0}$  is what one  normally thinks of as the Hamiltonian, and from which one determines the evolution in quantum theory. However, here we cannot  easily utilize ${\cal P}_0$ for the latter purpose,  because $x^0$ gets promoted to  a noncommuting operator in the quantum theory, which in particular will not commute with ${\cal  P}_0$.} 
 The quantum analogue of (\ref{Hameq}) gives a meaningful Heisenberg equation, because $\lambda$   remains a commuting parameter upon quantization.
Concerning the other Poincar\'e generators,  the action of the three-momenta  and Lorentz boosts generators on space-time is nonlinear, while the action of the angular momentum is undeformed.  The latter follows from 
\be  \{x_i,L_j\}_{\mbox{\tiny DB}} =\epsilon_{ijk}x_k\qquad\quad\{x_0,L_i\}_{\mbox{\tiny DB}}=0\label{LxDB}  \ee

Unlike with the standard gauge, here the position operators have discrete spectra  in the quantum theory.  This follows  since the subalgebra spanned by the spatial components of position and momenta coincides with that of Snyder.\cite{Snyder:1946qz}
More explicitly, define  
\be A_i =\frac 12(L_i +\Lambda x_i)\qquad\quad B_i =\frac 12(L_i -\Lambda x_i)\;,\ee
which from (\ref{poincarealgjj}), (\ref{xixjdb}) and (\ref{LxDB}), satisfy  two $SU(2)$ algebras 
\be \{A_i,A_j\}_{\mbox{\tiny DB}} =\epsilon_{ijk}A_k\qquad\quad
   \{B_i,B_j\}_{\mbox{\tiny DB}} =\epsilon_{ijk}B_k\qquad\quad \{A_i,B_j\}_{\mbox{\tiny DB}} =0   \ee  
   From the  $ x_i L_i  =0 $, it follows that 
$  A_i  A_i= B_i B_i$. 
  In the  quantum theory, we replace $A_i$ and $B_i$ by operators $\hat A_i$ and $\hat B_i$, and Dirac brackets 
   by commutators of the  operators divided by $i\hbar$.   Then $ \hat A_i \hat A_i$,  $\hat A_3$ and  $\hat B_3$ form a complete set of independent commuting operators.   
 These operators, along with  the coordinate  and angular momentum operators,  $\hat x_i$ and $\hat L_i$,  respectively, have discrete spectra. 
      Denoting the eigenvectors by $|j,m_A,m_B>$, $\; m_A, m_B=-j,1-j,...,j$, one has 
      \beqa
\hat A_i\hat A_i\;|j,m_A,m_B> &=&\hbar^2 j(j+1)\;|j,m_A,m_B>\cr & &\cr
\hat A_3\;|j,m_A,m_B> &=&\hbar\; m_A\;|j,m_A,m_B>\cr & &\cr
\hat B_3\;|j,m_A,m_B> &=&\hbar\; m_B\;|j,m_A,m_B> \;, \eeqa
   and also
\beqa
\hat x_3\;|j,m_A,m_B> &=&\frac \hbar{\Lambda} \;(m_A-m_B)\;|j,m_A,m_B>\cr & &\cr
\hat L_3\;|j,m_A,m_B> &=&\hbar\; (m_A+m_B)\;|j,m_A,m_B> 
\eeqa 
While $j,m_A$ and $m_B$ can be integers or half-integers, the eigenvalues of $\hat L_i$ are integers (times $\hbar$).  Representing $\hat L_i$ as the differntial operators $-{i\hbar}  \epsilon_{ijk} p_j \frac{\partial}{\partial p_k }\;,$ one then gets  singlevalued angular momentum eigenfunctions in momentum space.  The eigenvalues of $\hat x_i$ are integers times $\hbar/\Lambda$.

Additional remarks are: 

\noindent {\it i)} The eigenvectors $|j,m_A,m_B>$ are not stationary.  Neither are those which simultaneously diagonalize   $\hat L_i\hat L_i$,  $\hat L_3$, and  $\hat A_i\hat A_i$, since the last  operator does not commute with the Hamiltonian operator $\hat p^0$. 

\noindent {\it  ii)} Using the commutation relations
 \beqa [\hat A_i,\hat p_j]   &=& \frac{i\hbar}2\Bigl(\epsilon_{ijk} \hat p_k+\Lambda \delta_{ij} +\frac{\hat p_i\hat p_j}{\Lambda}\Bigl)\cr & &\cr [\hat B_i,\hat p_j]   &=& \frac{i\hbar}2\Bigl(\epsilon_{ijk} \hat p_k-\Lambda \delta_{ij} -\frac{\hat p_i\hat p_j}{\Lambda}\Bigl) \;,\eeqa one can write down
the following differential representation  for the $SU(2)$ generators on the space of square-integrable functions $\{{\cal F}(\vec p)\}$
 \beqa \hat A_i&=& \;\;\frac{i\hbar}2 \biggl(\Lambda \frac{\partial}{\partial p_i } +\frac  {p_i p_j}\Lambda \frac{\partial}{\partial p_j } - \epsilon_{ijk} p_j \frac{\partial}{\partial p_k }\biggr) \cr &&\cr
\hat  B_i&=& -\frac{i\hbar}2 \biggl(\Lambda \frac{\partial}{\partial p_i } +\frac { p_i p_j }\Lambda\frac{\partial}{\partial p_j } + \epsilon_{ijk} p_j \frac{\partial}{\partial p_k }\biggr)  \eeqa
The operators are symmetric for the scalar product defined using the measure $d^3p /(\Lambda^2 + \vec p^2)^2$.

\noindent {\it  iii)}  An alternative derivation of the Dirac brackets (\ref{xixjdb}-\ref{DBof xipj}) may be possible starting from a gauge fixed Lagrangian.  A first order formalism, analogous to \cite{Ghosh:2005rz},\cite{Banerjee:2006wf}, should be convenient for this purpose since the gauge constraint involves momentum variables.   The gauged fixed Lagrangian should then reduce to $-m\sqrt{1-\dot x_i \dot x_i}$, or its equivalent, in the limit $\Lambda\rightarrow \infty$.

\bigskip
\section{Discrete time gauge}
\setcounter{equation}{0}

Now we introduce a gauge condition  which from (\ref{fpa}) leads to a discrete spectrum for the time. Here, we shall assume that  one of the spatial coordinates $x^i$  can be identified with an  angular variable $\phi$,  $ 0\le \phi< 2 \pi$.  This implies the existence of either a coordinate singularity or a physical singularity on the spatial manifold.  The latter can be associated with a black hole with  axial symmetry. Let us assume this to be the case.  Then we then need to replace the Minkowski metric of the previous two sections with the appropriate black hole metric tensor $g_{\mu\nu}$.  $\phi$ corresponds to a Killing direction and its conjugate momenta $p_\phi$  is a constant along the particle geodesic. 
 If, furthermore, we assume the  metric tensor to be stationary, then there is an additional constant   $p_0$ conjugate to the time $x^0$.  Thus,
\be \{p_0,\Psi_1\}=\{p_\phi,\Psi_1\}=0\ee
Other momenta may not have vanishing Poisson brackets with $\Psi_1$ since the constraint  is a function of the background metric.

 For the gauge constraint $\Psi_2$ we now choose
 \be \Psi_2 = x^0 +\Theta p_\phi - \lambda\;\approx 0\label{dtgfxng} \;,\ee where $\Theta$ is a constant.  A similar choice was made in \cite{Pinzul:2004qu}.
 Substituting into (\ref{DB}) leads to nonvanishing Dirac brackets of the space-time coordinates with $\phi$,
 \be
 \{x^\mu,\phi\}_{\mbox{\tiny DB}}= \Theta \frac{p^\mu}{p^0}    \ee
 The Dirac brackets simplify if we replace  $\phi$ by
 \be \phi' = \phi +\Theta p_0  \;, \ee corresponding to a constant translation of the angular variable.  In terms of $\phi'$,  the only nonvanishing Dirac bracket between the space-time coordinates is  \be
 \{x^0,e^{i\phi'}\}_{\mbox{\tiny DB}}=i\Theta  e^{i\phi'} \ee 
 This agrees with the Poisson brackets of \cite{Chaichian:2001nw},\cite{Balachandran:2004yh}, and with  a special case ($c_2=-c_3$) of the brackets found in \cite{Dolan:2006hv} for the BTZ black hole, which preserves the isometry of the solution.  For the Dirac brackets of the momenta, one finds
\beqa 
\{x^\mu,p_i\}_{\mbox{\tiny DB}}&=&\delta^\mu_i +\frac \Theta{2 p^0}\;\delta^\mu_\phi \;\partial_i g^{\rho\sigma} p_\rho p_\sigma\cr& &\cr
\{x^\mu,p_0\}_{\mbox{\tiny DB}}&=& \delta^\mu_0 -\frac{p^\mu}{p^0} \cr& &\cr
\{p_i,p_0\}_{\mbox{\tiny DB}}&=&\frac 1{2 p^0}\; \partial_i g^{\mu\nu} p_\mu p_\nu\cr
& &\cr \{p_i,p_j\}_{\mbox{\tiny DB}} &=&0 \;, \eeqa 
where $g^{\mu\nu}$ denotes the inverse metric tensor and $\partial_i=\partial/\partial x^i$.

 In  the quantum theory,
  the operator analogues $\hat x^0$ and $e^{i\hat \phi'}$ of $ x^0$ and $e^{i\phi'}$, respectively, satisfy the commutation relation  
\be [\hat x^0,e^{i\hat \phi'}] = -\hbar \Theta e^{i\hat \phi'}\label{qntmalg} \ee
It follows that  $\exp\Bigl\{{-\frac{2\pi i\hat
  x^0}{\hbar \Theta}}\Bigr\}$ is a central element and  one can  identify it with   $e^{i\chi}\BI$ in an
irreducible representation of  the algebra. 
The spectrum for the time operator $\hat x^0$ is then discrete\footnote{For another derivation of a discrete time spectrum starting from  point particle dynamics in $2+1$ space-time, see \cite{Matschull:1997du}.}
 \be \hbar\Theta\Bigl(n
-\frac {\chi}{2\pi}\Bigr)\;,\qquad n\in
{\mathbb{Z}}\label{dsctsptmt}\ee    Implications of such a discrete time spectrum have  been discussed in  
 \cite{Chaichian:2001nw},\cite{Balachandran:2004yh}.

\subsubsection*{Acknowledgements}
I am very grateful for discussions with  Brian Dolan.

\end{document}